\documentclass[aps,prb,twocolumn,superscriptaddress]{revtex4-2}

\usepackage{amsmath}
\usepackage{amssymb}
\usepackage{graphicx}	% required for figures
\usepackage{comment}  % required for block-wise comment
\usepackage{bm}

\usepackage{todonotes}

\usepackage[colorlinks=true]{hyperref}	% use for hypertext links, including those to external documents and URLs

\begin{document}

    \title{Robust quantum anomalous Hall effect with spatially uncorrelated disorder}
	
	%Authors of paper
	\author{Kristof Moors}
        \thanks{Present Address: Imec, Kapeldreef 75, 3001 Leuven, Belgium}
	\email[]{kristof.moors@imec.be}
	\affiliation{Peter Grünberg Institute (PGI-9), Forschungszentrum Jülich, 52425 Jülich, Germany}
    \affiliation{JARA-Fundamentals of Future Information Technology, Jülich-Aachen Research Alliance, Forschungszentrum Jülich and RWTH Aachen University, 52425 Jülich, Germany}
	\author{Gen Yin}
    \email[]{gen.yin@georgetown.edu}
	\affiliation{Department of Physics, Georgetown University, Washington, D.C.~20057, USA}
	
	\date{\today}
	
	\begin{abstract}
            In magnetic topological insulators, a phase transition between a quantum anomalous Hall (QAH) and an Anderson localization phase can be triggered by the rotation of an applied magnetic field.
            Without the scattering paths along magnetic domains, this phase transition is governed by scattering induced by nonmagnetic disorder.
            We show that the QAH phase is strikingly robust in the presence of spatially uncorrelated disorder.
            The robustness is attributed to the quantum confinement induced by the short correlation length of the disorder.
            The critical behavior near the phase transition suggests a picture distinct from quantum percolation.
            This provides new insights on the robustness of the QAH effect in magnetic topological insulators with atomic defects, impurities, and dopants.
	\end{abstract}
	
	%\keywords{}
	
	\maketitle

\section{Introduction}
\label{sec:introduction}
Magnetic topological insulator (MTI) thin films have received a lot of attention as a host of the quantum anomalous Hall (QAH) effect~\cite{Yu2010, Chang2013, Tokura2019, Deng2020}.
This stems from the dissipationless transport of a topologically protected edge channel for which backscattering is strictly forbidden.
This edge channel is promising for various applications due to its quantum coherent nature, especially as an important building block of topological superconductors hosting chiral Majorana modes~\cite{Wang2013}.
An ideal QAH insulator (QAHI) system has the Fermi level inside the QAHI gap across the whole sample, but energy fluctuations due to disorder are inevitable in real systems.
For example, in magnetically modulated $\textrm{(Bi,Sb)}_2\textrm{Te}_3$, the doping with Sb atoms is essential, dragging the Fermi level down to the center of the bulk band gap~\cite{Chang2013,Kou2015,Checkelsky2014}.
Magnetic dopants such as Cr, Mn and V are usually added to replace the Bi sites, offering the long-range magnetic order that opens up a topological band gap near the Dirac point of the surface states~\cite{Mogi2017,Liu2018}.
Characterized by scanning tunneling microscopy, the spatial correlations of the disorder are typically at the atomic scale or up to a few nanometers~\cite{Lee2015, Chong2020, Park2024, Brede2024} and the energy fluctuation can be up to $\sim\textrm{0.1\thinspace\textrm{eV}}$~\cite{Chen2015}.
Surprisingly, the QAH effect can still emerge at low temperatures ($\sim 1\thinspace\textrm{K}$), and the quantization is robust and accurate.
A typical measurement of the QAH effect is performed by scanning the out-of-plane magnetic field.
When lowering the field, the magnetization originally along the out-of-plane easy axis shrinks, which lowers the topological band gap and can eventually drive the system into an Anderson localized insulating (ALI) state, yielding a quantum phase transition.
Several theoretical works have already studied the properties of this transition in MTI thin films with magnetic and nonmagnetic disorder~\cite{Onoda2003, Nomura2011, Chang2016, Qiao2016}.
Interestingly, this transition is similar to plateau transitions in quantum Hall systems~\cite{QHE1990, Huckestein1995}, where the critical behavior is known to match quantum percolation models~\cite{Trugman1983, Chalker1988}.
In recent experimental studies, the value of the scaling exponent has been reported to vary in a range ($\nu \approx 2.7$ in Ref.~\cite{Kawamura2020}, $\nu \approx 1.6$ in Ref.~\cite{Deng2022}), suggesting the rich underlying physics.
Recent theoretical studies have also pointed out that the domain walls formed near the coercive field can host extra topological edge modes, resulting in a Berezinskii-Kosterlitz-Thouless-type transition in the presence of structural inversion asymmetry~\cite{Chen2019}.
In this work, we closely examine the quantum phase transition in MTI thin films induced by a rotation of the external magnetic field, rather than a linear scan~\cite{Kawamura2018, Kawamura2020, Luan2023}.
With this approach, the magnetization can be dragged along the external magnetic field, avoiding the formation of magnetic domains.
The topological band gap eventually closes when the magnetization is completely in-plane.
Although the scattering paths provided by magnetic domains are absent in this setup, nonmagnetic disorder can still drive the system into an ALI phase~\cite{Checkelsky2014, Feng2015, Kou2015, Chang2016, Kawamura2018, Kawamura2020, Deng2022, Luan2023, Deng2023}.
Here, we show that the critical behavior of the phase transition strongly depends on the correlation length of the nonmagnetic disorder.
In particular, when the spatial correlation length approaches the nanometer scale, the critical exponent deviates from the quantum percolation value ($\nu \approx 2.6$) and varies sensitively with the disorder strength.
More importantly, the short correlation length preserves dissipationless transport even when the energy fluctuations significantly exceed the band gap of the clean limit, resulting in a quantized Hall conductivity that is unexpectedly robust.

\begin{figure*}[htb]
    \centering
    \includegraphics[width=\linewidth]{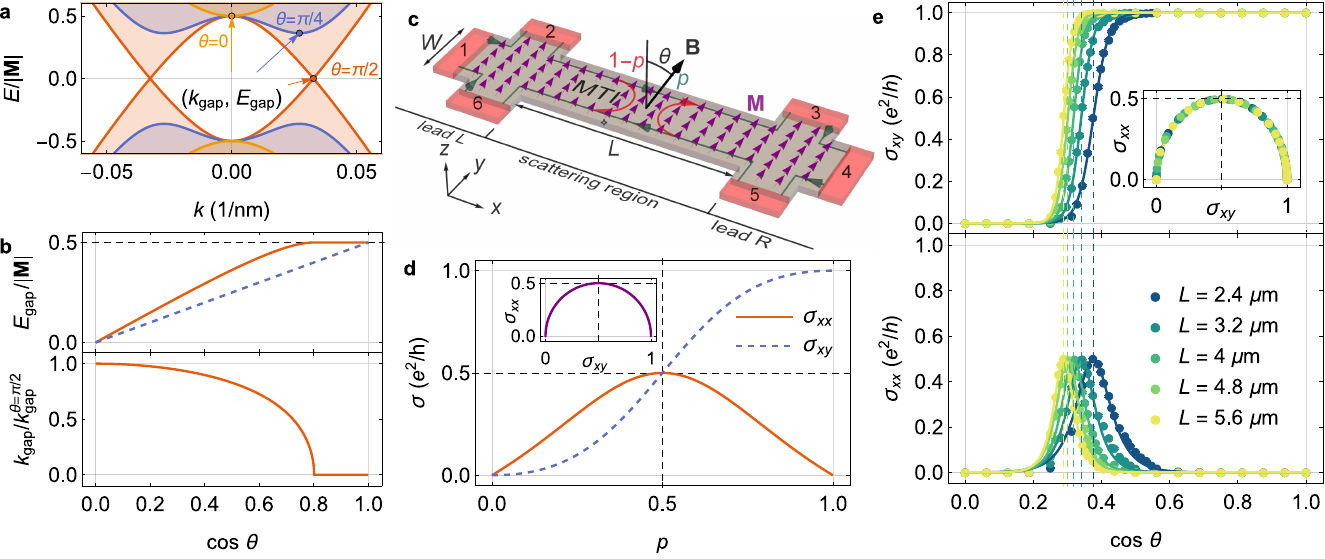}
    \caption{
        \textbf{QAH-to-trivial insulator transition driven by magnetization rotation in a MTI thin film.}
        \textbf{(a)} The spectrum of a MTI ribbon along $k_x$ for different orientation angles $\theta$ of the magnetization in the $y\text{-}z$ plane. The spectral gap closes when the magnetization is directed along $y$ ($\theta = \pi/2$).
        \textbf{(b)} The size and position in reciprocal space of the spectral gap minimum as a function of $\cos\theta$.
        \textbf{(c)} Schematic of a MTI thin-film Hall bar in the single-channel regime with rotating magnetization angle $\theta$ and transmission probability $p$ for a chiral edge channel across the central section of the Hall bar.
        This central section is the scattering region of the effective two-terminal setup of our microscopic quantum transport simulation approach (see text for details).
        \textbf{(d)} The longitudinal and Hall conductivities of the MTI Hall bar as a function of the transmission probability $p$, satisfying a semicircle relation (see inset).
        \textbf{(e)} The longitudinal and Hall conductivities for disordered MTI thin-film Hall bars ($E_\mathrm{F} = 0.2 \, E_\mathrm{gap}^{\theta=0}$, spatially uncorrelated disorder with  $S_\mathrm{dis} = 3 \, E_\mathrm{gap}^{\theta=0}$, and $y\text{-}z$ magnetization rotation plane) of different sizes (with identical aspect ratio $L/W = 80$) as a function of $\cos\theta$, based on microscopic quantum transport simulations. The transition value $\cos\theta^\ast$ is indicated for the different ribbon sizes with a vertical dashed line.
    }
    \label{fig:1}
\end{figure*}

The paper is structured as follows. In Sec.~\ref{sec:model}, we present our modeling and simulation approach. Results on the robustness of the QAHI phase and the finite-size scaling analysis of the QAH-to-trivial insulator transition are presented in Sec.~\ref{sec:results}, followed by the conclusion in Sec.~\ref{sec:conclusion}. More technical details and supporting results are presented in Appendices~\ref{app:bulk-spectrum} to \ref{app:size-scaling-alt}.

\section{Model}
\label{sec:model}
We consider a two-dimensional four-orbital tight-binding model on a square lattice.
This model is obtained from standard discretization of a four-band model of an MTI thin film~\cite{Yu2010,Wang2013} with a lattice constant $a=1\thinspace\textrm{nm}$:
\begin{equation} \label{eq:Hamiltonian}
\begin{split}
    H_\textsc{MTI}(k_x, k_y) &= \hbar v_\textsc{D} (k_y \sigma_x - k_x \sigma_z) \tau_z \\
    &\quad + [m_0 - m_1 (k_x^2 + k_y^2)] \tau_x + \mathbf{M} \cdot \bm{\sigma},
\end{split}
\end{equation}
where $\sigma_i$ and $\tau_i$ ($i=x,y,z$) are the Pauli matrices acting on the spin up/down and top/bottom surface subspaces, respectively.
The first term is the typical Rashba-Dirac term of TI surface states with Dirac velocity $v_\textsc{D}$.
The bulk spectrum of the MTI is not included in the continuum model, but the coupling of the top and bottom surfaces through the bulk is governed by parameters $m_0$ and $m_1$.
The magnetization of the material is represented by the vector $\mathbf{M}$.
In our quantum transport simulations, we consider the following parameters: $v_\textsc{D} = 3.5\times10^5$ m/s, $m_0 = 50$ meV, $m_1 = 15$ meV$\cdot$\r{A}$^2$ and $|\mathbf{M}| = 100$ meV.
With an out-of-plane orientation of the magnetization, the system is a QAHI with a gap $2 E_\mathrm{gap}^{\theta=0} = 2||\mathbf{M}| - |m_0|| = 100\,\textnormal{meV}$.

In this work, we consider a uniform (single-domain) magnetization across the MTI thin-film system with a fixed orientation and angle $\theta$ with respect to the $z$ axis: $M_z = |\mathbf{M}| \cos\theta$.
When the magnetization is rotated from out-of-plane to in-plane, the QAHI gap reduces and eventually closes (see Figs.~\ref{fig:1}a,b and Appendix~\ref{app:bulk-spectrum}).
We restrict our setup to nonmagnetic disorder (i.e., spectral fluctuations due to, e.g., charged impurities or the electrostatic environment), which is introduced as fluctuations of the onsite energies $\delta E_\mathrm{onsite}$.
Disorder is described by a Gaussian distribution, $f(\delta E) = 1/\sqrt{2\pi S_\mathrm{dis}^2}) \exp[-\delta E^2/(2 S_\mathrm{dis}^2)]$ with disorder strength $S_\mathrm{dis}$.
For uncorrelated disorder, this distribution is applied independently on each lattice site. For spatially correlated disorder, the following autocovariance function is considered:
\begin{equation}
    \langle \delta E(\mathbf{r}) \delta E(\mathbf{r}') \rangle = S_\mathrm{dis}^2 \exp[-(\mathbf{r} - \mathbf{r}')^2/(2\Lambda^2)],
\end{equation}
with correlation length $\Lambda$.
While disorder in lattice models of topological materials is commonly generated without spatial correlations~\cite{Onoda2003, Li2009, Groth2009, Chang2016, Qiao2016}, we find that these correlations can be of crucial importance with respect to the robustness of the QAH effect.

To capture the QAH effect in the modeled system, we perform quantum transport simulations implemented in Kwant~\cite{Groth2014}, with the parallel sparse direct solver MUMPS~\cite{Amestoy2001}.
\begin{figure}
    \centering
    \includegraphics[width=.98\linewidth]{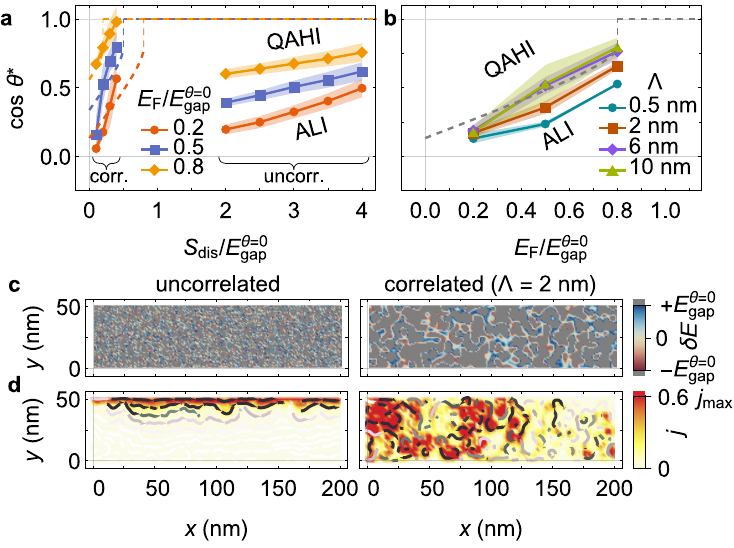}
    \caption{
        \textbf{Robustness of QAH effect with respect to correlated and uncorrelated disorder.}
        \textbf{(a)} The transition values of $\cos\theta^\ast$ as function of disorder strength when rotating the magnetization in the $y\text{-}z$ plane ($W = 0.05\,\text{\textmu m}$, $L = 4\,\text{\textmu m}$). The shaded areas indicate the full width at half maximum of the $\sigma_{xx}$ curve (see examples in the lower panel of Fig.~\ref{fig:1}e). The dashed lines indicate the expected phase boundaries based solely on conventional band-smearing picture. The data points for correlated disorder are obtained for $\Lambda = 10\,\text{nm}$.
        \textbf{(b)} Values of $\cos\theta^*$ as a function of the Fermi energy for different correlation lengths. Here, the disorder strength is fixed at $S_\mathrm{dis} = 0.2 \, E_\mathrm{gap}^{\theta=0}$. The dashed lines and the shaded areas are similar to those in (a).
        \textbf{(c)} An example of uncorrelated (left) and correlated (right) disorder completely smearing the topological band gap. The red–blue color scale indicates local fluctuations $\delta E$ with Fermi level remaining within the gap, whereas the gray color indicates local fluctuations exceeding the gap size.
        \textbf{(d)} The current density of a QAH edge state injected from the left in to the disorder profiles shown in (c). In both panels, we have $E_\mathrm{F} = 0.2 \, E_\mathrm{gap}^{\theta=0}$, $\cos\theta = 0.3$ ($y\text{-}z$ rotation) and $S_\mathrm{dis} = 2 \, E_\mathrm{gap}^{\theta=0}$.
    }
    \label{fig:2}
\end{figure}
{For studying the QAH-to-trivial insulator transition, we consider a MTI Hall bar and analyze the longitudinal and Hall conductivities. To reduce the computational burden of the quantum transport simulations and allow for extensive disorder averaging}, we only consider disorder in the central section of the MTI {Hall bar} between the left (1,2,6) and right (3,4,5) terminals (see Fig.~\ref{fig:1}c) such that the 6-terminal setup can be reduced to a two-terminal transport picture{, with leads 1,2,6 (3,4,5) effectively forming a single left (right) lead (see Fig.~\ref{fig:1}c)}.
The transmission probability $p$ over the central section {can then be} extracted from a two-terminal scattering matrix, {with the} longitudinal and Hall conductivities given by: $\sigma_{xx} = \sigma_0 (p - p^2)/[1 + 2 p(p - 1)]$ and $\sigma_{xy} = \sigma_0 p^2/[1 + 2p(p - 1)]$, respectively (see Fig.~\ref{fig:1}d), where $\sigma_0 \equiv e^2/h$ is the quantized sheet conductivity.
With $p$ ranging between 0 and 1, the extracted conductivities cover the QAH ($p=1$, $\sigma_{xx}=0$, $\sigma_{xy}=\sigma_0$) and trivial ($p=\sigma_{xx}=\sigma_{xy}=0$) insulator phases, as well as the transition between them, which satisfies the semicircle law~\cite{Pruisken1988, Dykhne1994, Hilke1999}: $(\sigma_{xx}/\sigma_0)^2 + (\sigma_{xy}/\sigma_0 - 1/2)^2 = (1/2)^2$.
This can be seen from the inset of Fig.~\ref{fig:1}d.
Details {on} this {effective two-terminal} setup can be found in Appendix~\ref{app:conductance-matrix}.
For our analysis, we employ the transfer matrix method to extend the length of the central section beyond the length of the simulation domain, as described in Appendix~\ref{app:transfer-matrix}.
Furthermore, we consider disorder averaging. Instead of doing a single simulation, we perform such simulations on $N_\mathrm{dis} = 5000$ systems with different landscapes of disorder, and then extract the average transmission probability (see Appendix~\ref{app:disorder-averaging}).
To understand the critical behavior near the phase transition, one needs to find the parameters that cover both the quantized and the localized plateau.
One set of such transition curves can be seen in Fig.~\ref{fig:1}e, where the Hall conductivity serves as the order parameter when we rotate the magnetization at an angle $\theta$ away from the $z$ axis: $\cos\theta \in [0, 1]$.
The transition curves are well described by a hyperbolic tangent profile, as shown by the lines through the data points in Fig.~\ref{fig:1}e.
We use two fitting parameters to characterize the position of the transition, i.e., the transition value $\cos\theta^\ast$ with $\sigma_{xx}(\cos\theta^\ast) = \sigma_{xy}(\cos\theta^\ast) = e^2/(2h)$, and the steepness of the transition, respectively (see Appendix~\ref{app:transition-curves}).

\section{Results}
\label{sec:results}

\subsection{Robustness of quantum anomalous Hall phase}
\label{subsec:robustness}
\begin{figure}
    \centering
    \includegraphics[width=\linewidth]{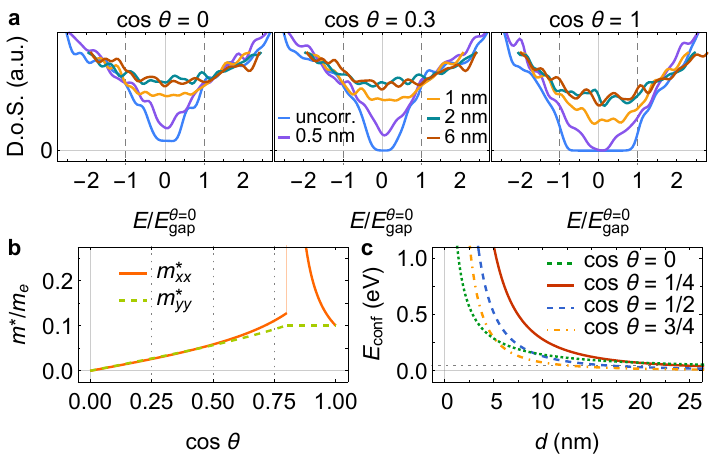}
    \caption{
        \textbf{Robustness of spectral gap.}
        \textbf{(a)} The bulk density of states for a MTI thin film at different magnetization orientations and disorder correlation lengths ($\Lambda$). The disorder strength is set at $S_\mathrm{dis} = 2 \, E_\mathrm{gap}^{\theta=0}$ for all three panels.
        \textbf{(b)} The effective mass along $x$ and $y$ at the bulk band edges (indicated by the arrows in Fig.~\ref{fig:1}a).
        \textbf{(c)} The confinement quantization energy of a bulk state as a function of the disorder puddle size $d$ for different magnetization rotation angles.
        These confinement energies drop back below the band edge energy $E_\mathrm{gap}^{\theta=0}$ (horizontal gray dashed line) for large $d$.
    }
    \label{fig:3}
\end{figure}

The transition between the QAHI and ALI phases is sensitive to both the Fermi level position and the disorder characteristics.
We demonstrate this by extracting the transition values $\cos\theta^\ast$ of the phase transition for different system parameters.
In Fig.~\ref{fig:2}, we present $\cos\theta^\ast$ for spatially correlated and uncorrelated disorder as a function of the Fermi level position (\ref{fig:2}a) and disorder strength (\ref{fig:2}b), while keeping other parameters identical.
As expected, the QAHI phase withstands larger out-of-plane orientation angles for a Fermi level that is more centrally positioned in the QAHI gap (i.e., closer to the Dirac point) and for lower disorder strengths.
The QAHI-ALI phase transition is expected to occur when the edge state is able to backscatter into the opposing counterpropagating state across the bulk, through a combination of scattering via bulk states and quantum tunneling~\cite{Chalker1988}.
The transition can be expected near the condition $\mu \pm S_\mathrm{dis} \pm E_\mathrm{gap}(\cos\theta) = 0$, where the disorder strength aligns the local Fermi level with the band edge of the bulk spectrum above or below the QAHI gap.
This condition is indicated by the dashed curves in Figs.~\ref{fig:2}a-b.
In both figures, the area above the dashed curves is expected to be the QAHI phase and the other side should be the ALI phase.
Note that this expected phase boundary does not depend on the correlation length of disorder, as it follows solely from a comparison of energy scales.
When the disorder is correlated over several nms, the phase boundary is indeed well captured by our numerical results, as shown by the points on the left of Fig.~\ref{fig:2}a and those in Fig.~\ref{fig:2}b.
Surprisingly, the phase diagram captured by our numerical results is significantly distinct from the conventional understanding (i.e., the phase boundary based on energetic considerations) when the disorder approaches the nm scale or is fully uncorrelated.
The QAHI phase appears to be much more robust against spectral fluctuations and still survives even when the disorder strength is well above the expected percolation threshold {(i.e., $S_\mathrm{dis} = \text{min}_\pm\{|E_\mathrm{gap}(\cos\theta) \pm \mu|\}$)}.
This is represented by the {appearance} of the QAHI phase {for a finite range of $\cos\theta$ values, as indicated} by the points on the right of Fig. \ref{fig:2}a.
Two example profiles of the local chemical potential due to the nonmagnetic disorder are shown in Fig.~\ref{fig:2}c, where the gray (red-white-blue) color scale indicates local fluctuations that (do not) exceed the bulk topological gap.
The large gray area in Fig.~\ref{fig:2}c suggests that significant percolation is expected for both the uncorrelated (left) and correlated (right) cases.
Notably, the edge state can almost transmit perfectly through the uncorrelated disorder, whereas severe percolation and backscattering are captured in the case of correlated disorder ($\Lambda=2\thinspace\text{nm}$).
The extra robustness of the QAH effect stems from the insensitivity of the bulk spectrum against uncorrelated disorder.
To show this, we examine the density of states (DOS) of a simulation domain of $0.2 \times 0.05 \,\text{\textmu m}^2$, considering periodic boundary conditions so that the topological edge states do not show up (see Fig.~\ref{fig:3}a).
The DOS is resolved with the kernel polynomial method~\cite{Weisse2006} with a resolution of $2.5\,\text{meV}$.
The three panels correspond to different rotation angles of the magnetization.
In all three cases, the disorder strength $S_\textrm{dis}=2 \, E^{\theta=0}_\textrm{gap}$ exceeds the bulk topological gap by a factor of two.
Generally, the topological gap closes in all three cases, as expected.
However, the gap starts to restore when the correlation length of the disorder approaches the nanometer scale and is almost fully restored when the disorder is uncorrelated.

\begin{figure}[tb]
    \centering
    \includegraphics[width=0.95\linewidth]{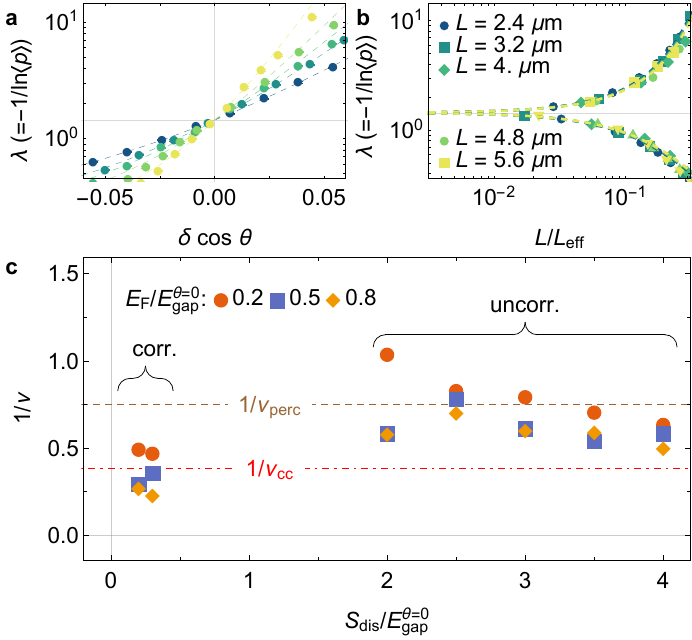}
    \caption{
        \textbf{Finite-size scaling analysis of QAHI-ALI phase transition.}
        \textbf{(a)} The localization length $\lambda$ as a function of $\delta\!\cos\theta \equiv \cos\theta - \cos\theta^\ast$ for different MTI ribbon lengths.
        \textbf{(b)} Collapsed values of $\lambda$ as a function of $L/L_\textrm{eff}$ using the extracted exponent for the effective size: $L_\mathrm{eff}(\cos\theta)=A |\delta\!\cos\theta|^{-\nu}$, with $A=1\,\text{\textmu m}$ here without loss of generality. We consider $E_\mathrm{F} = 0.2 \, E_\mathrm{gap}^{\theta=0}$, $S_\mathrm{dis} = 2\,E_\mathrm{gap}^{\theta=0}$ (uncorrelated), and $x\text{-}z$ rotation in both (a) and (b).
        \textbf{(c)} The scaling exponent $1/\nu$ as a function of disorder strength for spatially correlated (left, $\Lambda = 10\,\text{nm}$) and uncorrelated (right) disorder with different Fermi levels, considering $x\text{-}z$ rotation. The values of classical percolation $1/\nu_\textrm{perc}$ and the Chalker–Coddington value $1/\nu_\textrm{CC}$ are indicated by horizontal dashed lines for reference.
    }
    \label{fig:4}
\end{figure}

The resilience of the QAHI gap against uncorrelated disorder can be understood by considering the confinement energy of the bulk states.
At low energies, the bulk states are effectively trapped in regions $d \propto \Lambda$ due to disorder fluctuations (see, for example, the extent of the gray patches in Fig.~\ref{fig:2}c).
This gives rise to a confinement energy that pushes up the energy levels by $E_\mathrm{conf} \propto 1/d^2$, which is determined by the effective mass of the gap (or $\propto 1/d$ for massless states).
We can extract the effective mass of the surface states from the curvature at the minimum of the spectrum (see Fig.~\ref{fig:3}b).
The sudden jump near $\cos\theta = 0.8$ for $m_{xx}$ is due to the band-edge transition from the two valleys (blue in Fig.~\ref{fig:1}a) to a direct band gap at $\Gamma$.
In Fig.~\ref{fig:3}c, we roughly estimate the confinement energy for different patch sizes by considering hard-wall confinement along $x$ and $y$.
When the spectrum is massive (gapped), the confinement energy is given by the particle-in-a-box ground state, whereas the value for $\cos\theta = 0$ is given by the massless (gapless) solution $E_\mathrm{conf} \approx 2 \hbar v_\mathrm{D} (\pi/d)$.
It is clear that the confinement energy can easily exceed the QAHI gap when the patch size is only a few nanometers.
Furthermore, the confinement energy increases when the magnetization is rotated from out-of-plane to in-plane orientation, which enhances the confinement due to the reduction of the effective mass.

\subsection{Finite-size scaling}
\label{subsec:finite-size-scaling}
The critical behavior of the QAHI-ALI transition in the presence of uncorrelated disorder is fundamentally different from the case where the disorder is correlated (with $\Lambda$ in the range of several nms).
We demonstrate this by identifying different finite-size scaling behaviors of the transition.
For this analysis, we consider the MacKinnon-Kramer fitting procedure~\cite{MacKinnon1983} with dimensionless localization length $\lambda \equiv -1/\ln \langle p \rangle$, where $\langle p \rangle$ is the ensemble-averaged transmission probability.
In Fig.~\ref{fig:4}a, example values of $\lambda$ are shown as a function of $\delta\!\cos\theta \equiv \cos\theta - \cos\theta^\ast$ for different system sizes, all crossing through $\lambda_\mathrm{c} \equiv 1/\ln2$ at $\delta\!\cos\theta = 0$, as they should.
The numerical results agree well with a single-parameter scaling of the localization length, $\lambda(L, \cos\theta) = f[L/L_\mathrm{eff}(\cos\theta)]$, where $L_\mathrm{eff}(\cos\theta) \propto |\delta\!\cos\theta|^{-\nu}$ is the control parameter-dependent effective size that obeys a power-law relation with critical exponent $\nu$.
As can be seen in Fig.~\ref{fig:4}b, all the values of $\lambda(L, \cos\theta)$ collapse on the same curve on both sides near the transition, confirming the single-parameter scaling relation.
We implemented a fitting procedure that directly yields this effective-size scaling exponent $\nu$ (see Appendix~\ref{app:size-scaling} for details) for different values of the Fermi level, disorder strength, and correlation length.
The critical exponents are presented as a function of disorder strength in Fig.~\ref{fig:4}c and reference values for classical percolation ($\nu_\mathrm{perc} = 4/3$) and quantum percolation according to the Chalker-Coddington model~\cite{Chalker1988} ($\nu_\mathrm{CC} \approx 2.6$) are indicated by the horizontal dashed lines.
For correlated disorder ($\Lambda = 10\,\text{nm}$), the numerical exponent values reasonably agree with the Chalker-Coddington value.
In this case, the transition is captured for disorder strengths lower than the topological gap ($S_\textrm{dis}/E^{\theta=0}_\textrm{gap}<1$), which is consistent with the quantum percolation picture.
However, the exponent values display a pronounced dependence on the disorder strength (significantly exceeding the topological gap) when the disorder is uncorrelated.
This suggests that the confinement not only introduces extra robustness to the QAHI phase, but also modifies the quantum phase transition in a nonuniversal manner.
Only for the largest disorder strengths is there a trend towards universal Chalker-Coddington-like scaling.
Besides the MacKinnon-Krammer approach, we also extracted the effective-size scaling exponent directly from fitting a power law to the size dependence of the maximal slope of the transition curve, which has been used in several recent experiments~\cite{Kawamura2020,Deng2023}, $\max\{\mathrm{d}\sigma_{xy}/\mathrm{d}\cos\theta\} \propto L^{1/\nu}$.
The details of this approach are presented in Appendix~\ref{app:size-scaling-alt} where we find good agreement with the results extracted by our single-parameter fitting procedure, indicating good numerical stability.
Note that nonuniversal finite-size scaling is also retrieved for correlated disorder with very short correlating lengths, i.e., in the (sub-)nm range (see Appendix~\ref{app:size-scaling-alt}), which coincides with the type of disorder for which we also recover an enhanced robustness of the QAHI phase.
The universal behavior of the QAH-to-trivial insulator transition (or lack thereof) thus offers a straightforward indication of the correlation length of nonmagnetic disorder in QAH experiments, which can be tuned through various synthesis parameters, including substrate selection, doping levels, and dopant species.

\section{Conclusion}
\label{sec:conclusion}
We have investigated the magnetization rotation-driven QAHI-to-ALI phase transition in MTI thin films with nonmagnetic disorder.
Due to the confinement energy of low-energy bulk states that are effectively trapped in regions with size proportional to the disorder correlation length, the QAHI phase is strikingly resilient with respect to atomic-scale disorder.
The QAHI phase withstands energy fluctuations significantly overwhelming the QAHI gap, and the corresponding effective-size scaling exponent displays nonuniversal behavior.
Our results shed new light on the robustness of the QAH effect in MTI thin films with uniformly random disorder at the atomic scale.
This offers a new avenue in the search for ideal QAHI systems besides clean and pristine Chern insulators.

\begin{acknowledgments}
This paper is based upon work supported by the National Science Foundation (US) under Grant \# ECCS-2151809 (G.Y.) and DMR-2440337 (G.Y.).
The authors acknowledge the discussions with Dr. Peter Olmsted at Georgetown University. This work is also supported by the QuantERA grant MAGMA and by the German Research Foundation under grant 491798118.
K.M.\ further acknowledges the financial support by the Ministry of Economic Affairs, Regional Development and Energy within Bavaria’s High-Tech Agenda Project ``Bausteine für das Quantencomputing auf Basis topologischer Materialien mit experimentellen und theoretischen Ans\"atzen'' (Grant No.\ 07 02/686 58/1/21 1/22 2/23), and by the German Federal Ministry of Education and Research (BMBF) via the Quantum Future project ‘MajoranaChips’ (Grant No.\ 13N15264) within the funding program Photonic Research Germany.
\end{acknowledgments}

\appendix
\renewcommand{\thefigure}{A\arabic{figure}}
\setcounter{figure}{0}
\begin{widetext}
\section{Spectrum of MTI model Hamiltonian}
\label{app:bulk-spectrum}

The bulk spectrum of Eq.~\ref{eq:Hamiltonian} with magnetization oriented in the $y$-$z$ plane ($\mathbf{M} = |\mathbf{M}|(0, \sin\theta, \cos\theta)$) is given by:

\begin{align}
   E_{\mu,\nu}(k_x, k_y) &= \mu \left(\hbar^2 v_\mathrm{D}^2 k^2 + |\mathbf{M}|^2 + (m_0 - m_1 k^2)^2 + \nu |\mathbf{M}| \sqrt{4 (m_0 - m_1 k^2)^2 + 2 \hbar^2 v_\mathrm{D}^2 k_x^2 [1 - \cos(2\theta)]}\right)^{\frac{1}{2}},
\end{align}
\end{widetext}
with $k^2 \equiv k_x^2 + k_y^2$ and $\mu, \nu = \pm 1$.
The two lowest-energy bands ($\mu = \pm 1, \nu=-1$) are shown in Fig.~\ref{fig:1}a for different magnetization rotation angles $\theta$. With in-plane orientation ($\theta = \pi/2$), the gap closes at $k_x = \pm (2 m_0 m_1 - \hbar^2 v_\mathrm{D}^2 + \sqrt{\hbar^4 v_\mathrm{D}^4 + 4 m_1^2 |\mathbf{M}|^2} - 4 m_0 m_1 \hbar^2 v_\mathrm{D}^2)^{1/2}/\sqrt{2 m_1^2}$.
The phenomenology remains similar when the magnetization is rotated along another in-plane direction.

\section{Conductance matrix}
\label{app:conductance-matrix}
We consider the Landauer-B\"uttiker formula for multiterminal ballistic transport, $I_i = (e/h) \sum_j (T_{ij} \delta\mu_j - T_{ji} \delta\mu_i)$, with terminal indices $i,j$ and transmission coefficients $T_{ij}$.
The corresponding conductance matrix $G_{ij}$ links the terminal currents with arbitrary chemical potential differences: $I_i = \sum_j G_{ij} (\delta \mu_j/e)$, where $e>0$ is the elementary charge and $\delta \mu_i=\mu_i-\mu_\textrm{eq}$ denotes the chemical potential shift with respect to equilibrium in terminal $i$.
Enforcing current conservation, the conductance matrix can be written as
\begin{align}
    G_{ij} &= \frac{e^2}{h} \times \begin{cases}
        T_{ij} & (i \neq j) \\
        -\sum\limits_{k \neq i} T_{ki} & (i = j)
    \end{cases},
\end{align}
Using the Hall bar setup illustrated in Fig. \ref{fig:1}c, we assume $i,j \in \{1,\ldots, 6\}$, with $\delta\mu_1 = - \delta\mu_4 = \delta\mu/2$ and $I_{2,3,5,6} = 0$.
The linear-response Hall effect can be obtained as the solution of a linear system, as detailed below.
Considering the geometry of the Hall bar, the conductance matrix can be rearranged in longitudinal and transverse blocks,
\begin{equation}
    \begin{pmatrix}
        \mathbf{I}_{\parallel} \\[4pt]
        \mathbf{I}_{\perp}
    \end{pmatrix}
    =
    \begin{pmatrix}
        G_{\parallel\parallel} & G_{\parallel\perp} \\[4pt]
        G_{\perp\parallel} & G_{\perp\perp}
    \end{pmatrix}
    \begin{pmatrix}
        \bm{\delta\mu}_{\parallel}/e \\[4pt]
        \bm{\delta\mu}_{\perp}/e
    \end{pmatrix},
\end{equation}
where $\parallel\in\{1,4\}$ and $\perp\in\{2,3,5,6\}$. 
Here, we have
\begin{align}
    \bm{\delta\mu}_{\parallel} &=
    \begin{pmatrix}
        +\delta\mu/2 \\[3pt]
        -\delta\mu/2
    \end{pmatrix},
    &
    \mathbf{I}_{\parallel} &=
    \begin{pmatrix}
        I_1 \\[3pt]
        I_4 
    \end{pmatrix}, \\
    \mathbf{I}_{\perp} &= \begin{pmatrix}
        0 \\[3pt]
        0 \\[3pt]
        0 \\[3pt]
        0
    \end{pmatrix},
    &
    \bm{\delta\mu}_{\perp} &=
    \begin{pmatrix}
        \delta\mu_2 \\[3pt]
        \delta\mu_3 \\[3pt]
        \delta\mu_5 \\[3pt]
        \delta\mu_6
    \end{pmatrix}.
\end{align}
Solving for the unknown terminal potentials and the longitudinal currents, we obtain
\begin{align} \label{eq:delta-mu_perp}
    \bm{\delta\mu}_{\perp} &= -\,G_{\perp\perp}^{-1}\,G_{\perp\parallel}\, \bm{\delta\mu}_{\parallel}, \\[6pt] \label{eq:I_para}
    \mathbf{I}_{\parallel} &= \big(G_{\parallel\parallel} - G_{\parallel\perp}\,G_{\perp\perp}^{-1}\,G_{\perp\parallel}\big)\,
    \bm{\delta\mu}_{\parallel}/e.
\end{align}
The longitudinal and Hall resistivities can now be obtained as follows,
\begin{align}
    \rho_{xx} &= \frac{\delta \mu_3 - \delta\mu_2}{{e} I_1}, \\
    \rho_{xy} &= \frac{\delta\mu_2 - \delta\mu_6}{{e} I_1},
\end{align}
with $\delta\mu_2$, $\delta\mu_3$, $\delta\mu_6$, and $I_1$ the solutions of Eqs.~\ref{eq:delta-mu_perp} and \ref{eq:I_para}.
From these resistivities, we can also obtain the longitudinal ($\sigma_{xx}$) and Hall ($\sigma_{xy}$) conductivities:
\begin{align}
    \begin{split}
        \sigma_{xx} &= \frac{\rho_{xx}}{\rho_{xx}^2 + \rho_{xy}^2} \\
        &= \frac{(\delta\mu_3 - \delta\mu_2) e I_1}{(\delta\mu_2 - \delta\mu_6)^2 + (\delta\mu_3 - \delta\mu_2)^2},
    \end{split} \\
    \begin{split}
        \sigma_{xy} &= \frac{\rho_{xy}}{\rho_{xx}^2 + \rho_{xy}^2} \\
        &= \frac{(\delta\mu_2 - \delta\mu_6) e I_1}{(\delta\mu_2 - \delta\mu_6)^2 + (\delta\mu_3 - \delta\mu_2)^2}.
    \end{split}
\end{align}
Now, we simplify the 6-terminal transport picture into a 2-terminal one.
This makes it easier to numerically capture the critical behavior when $L$ is large.
Following the chiral convention of Fig.~\ref{fig:1}c, we assume that
\begin{align} \label{eq:T-values}
    T_{ij} &= \begin{cases}
        1 & (i,j)=(2,1),(4,3),(5,4),(1,6) \\
        p & (i,j)=(3,2), (6,5) \\
        1 - p & (i,j)=(6,2), (3,5) \\
        0 & \textnormal{else}
    \end{cases},
\end{align}
where $p$ is the two-terminal transmission probability. 
This is effectively considering a scattering region limited to the long middle section of the Hall bar, as shown in Fig.~\ref{fig:1}c.

Plugging these transmission probabilities in the equations above, we obtain the following conductivities as a function of $p$:
\begin{align} 
    \sigma_{xx} &= \frac{e^2}{h} \frac{p(1-p)}{p^2 + (1 - p)^2} \label{eq:sigma-p}, \\
    \sigma_{xy} &= \frac{e^2}{h} \frac{p^2}{p^2 + (1 - p)^2} \label{eq:sigma_xy-p}.
\end{align}
Although this simplification enforces an exact semicircle relation between $\sigma_{xx}$ and $\sigma_{xy}$ and neglects the scattering details near the transverse terminals (2, 3, 5, and 6), it enables the use of the transfer-matrix approach to simulate an extended disordered region, thereby capturing the critical behavior for large values of $L$.

\section{Transfer matrix and length extension}
\label{app:transfer-matrix}
To simulate transport through longer disordered regions, we consider the transfer matrix method to extend the system length in a modular fashion. 
For each Kwant simulation, we have the $2\times2$ scattering matrix for an MTI ribbon of length $L_0$, which is defined as
\begin{equation}
    S = \begin{pmatrix}
        s_{\textrm{L}\leftarrow\textrm{L}} & s_{\textrm{L}\leftarrow\textrm{R}} \\
        s_{\textrm{R}\leftarrow\textrm{L}} & s_{\textrm{R}\leftarrow\textrm{R}}
    \end{pmatrix},
\end{equation}
where the subscripts denote the scattering amplitude from incoming states in the left or right lead to outgoing states. 
These complex amplitudes depend on the specific microscopic disorder configuration of the system. The transmission probability through the segment is given by
\begin{equation}
    p = |s_{\textrm{L} \leftarrow \textrm{R}}|^2 = |s_{\textrm{R} \leftarrow \textrm{L}}|^2.
\end{equation}

The scattering matrix $S$ relates incoming states in the leads to the outgoing states via:
\begin{equation}
\begin{pmatrix}
    |o_\textrm{L}\rangle \\
    |o_\textrm{R}\rangle
\end{pmatrix}
=
S
\begin{pmatrix}
    |i_\textrm{L}\rangle \\
    |i_\textrm{R}\rangle
\end{pmatrix},
\end{equation}
where $|i_\textrm{L,R}\rangle$ and $|o_\textrm{L,R}\rangle$ denote incoming and outgoing propagating modes in the left and right leads, respectively.
To build up longer systems, we recast the problem in terms of the transfer matrix $M$, which connects the propagating modes at one end of the system to those at the other:
\begin{align}
    \begin{pmatrix}
        |i_\textrm{R}\rangle \\
        |o_\textrm{R}\rangle
    \end{pmatrix}
    =
    M
    \begin{pmatrix}
        |i_\textrm{L}\rangle \\
        |o_\textrm{L}\rangle
    \end{pmatrix},
\end{align}
where $M$ can be expressed in terms of the scattering matrix components as:
\begin{equation}
    M =
    \begin{pmatrix}
        -s_{\textrm{L}\leftarrow\textrm{L}} / s_{\textrm{L}\leftarrow\textrm{R}} & 1 / s_{\textrm{L}\leftarrow\textrm{R}} \\
        s_{\textrm{R}\leftarrow\textrm{L}} - s_{\textrm{L}\leftarrow\textrm{L}} s_{\textrm{R}\leftarrow\textrm{R}}/s_{\textrm{L}\leftarrow\textrm{R}} & s_{\textrm{R}\leftarrow\textrm{R}} / s_{\textrm{L}\leftarrow\textrm{R}}
    \end{pmatrix}.
\end{equation}

To construct a system of extended length $L=N_\textrm{ext} L_0$, we concatenate $N_\textrm{ext}$ such segments by multiplying their respective transfer matrices:
\begin{equation}
    M_\textrm{ext} = M_1 M_2 \cdots M_{N_\textrm{ext}}.
\end{equation}
The total transmission probability through the extended system is then given by: 
\begin{equation}
    p_\textrm{ext} = |[M_\textrm{ext}]_{2,1}|^2.
\end{equation}
\begin{figure}
    \centering
    \includegraphics[width=\linewidth]{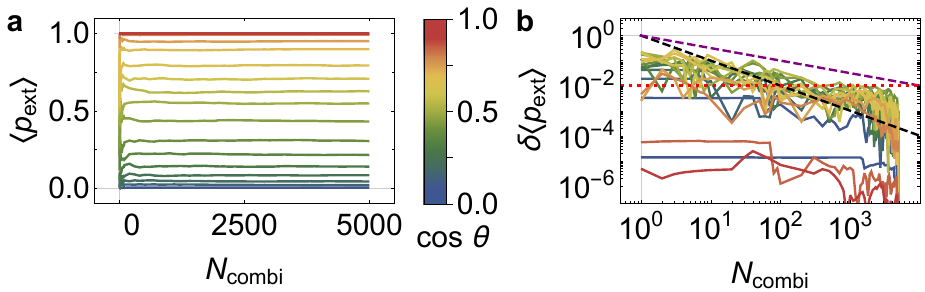}
    \caption{
        \textbf{Disorder-averaged transmission probability.}
        \textbf{(a)} The average transmission probability $\langle p_\textrm{ext} \rangle$ as a function of the number of combinations of disorder configurations $N_\textrm{combi}$ (see text for details) up to a maximum of $N_\textrm{combi} = 5000$ for different values of $\cos\theta$.
        \textbf{(b)} The difference of the mean with respect to the overall mean ($\delta \langle p_\textrm{ext} \rangle \equiv |\langle p_\textrm{ext} \rangle_{N_\textrm{combi}} - \langle p_\textrm{ext} \rangle_{5000}|$) as a function of the number of combinations $N_\textrm{combi}$ up to a maximum of $N_\textrm{combi} = 5000$ for different values of $\cos\theta$ on a log-log scale. The slope of $1/N_\textrm{combi}^{1/2}$ and $1/N_\textrm{combi}$ is shown for comparison with purple and black dashed lines, respectively.
        The transmission probabilities here are obtained for an MTI ribbon with $W = 0.05\,\textnormal{\textmu m}$, $L = 0.2 \, (4) \, \textnormal{\textmu m}$ ($N_\textrm{ext} = 20$), $E_\mathrm{F} = 0.2 \, E_\mathrm{gap}^{\theta=0}$, spatially uncorrelated disorder with $S_\mathrm{dis} = 3 \, E_\mathrm{gap}^{\theta=0}$, and magnetization rotated in the $x\text{-}z$ plane.
    }
    \label{fig:S1}
\end{figure}

\section{Disorder average}
\label{app:disorder-averaging}
We use the following disorder averaging procedure to compute the transmission probability $p_\textrm{ext}$ for a system of extended length $L=N_\textrm{ext}L_0$. 
First, we numerically compute the scattering matrix for a system of fixed length $L_0$ across $N_\textrm{dis}$ independent disorder realizations (with $N_\textrm{dis} > N_\textrm{ext}$). 
To construct a longer disordered system, we randomly select $N_\textrm{ext}$ distinct scattering matrices from this set, sequentially compose their corresponding transfer matrices, and extract the total transmission probability $p_\textrm{ext}$ of the extended system.

To obtain a reliable disorder average, we repeat this procedure for $N_\textrm{combi}$ independent combinations of $N_\textrm{ext}$ segments (with $N_\textrm{combi} \ll \binom{N_\mathrm{dis}}{N_\textrm{ext}}$ ), and compute the average and standard deviation of the resulting \( p_\textrm{ext} \) values. As shown in Fig.~\ref{fig:S1}, the average exhibits the expected \( 1/\sqrt{N_\textrm{combi}} \) convergence to the mean with increasing \( N_\textrm{combi} \). For the results presented in this work, we use \( N_\textrm{ext} = 20 \), \( N_\mathrm{dis} = 5000 \), and \( N_\textrm{combi} = 5000 \), ensuring that the statistical uncertainty in \( \langle p_\textrm{ext} \rangle \) remains below 1\% (see red dotted line in Fig.~\ref{fig:S1}b).
Replacing $p$ in Eqs.~\ref{eq:sigma-p} and~\ref{eq:sigma_xy-p} by $\langle p_\textrm{ext}\rangle$, we obtain the disorder-averaged values for the longitudinal and Hall conductivities. 
Examples of such disorder-averaged values are illustrated in Fig.~\ref{fig:1}e for different values of $\cos\theta$ and $L$. 
For an Anderson-localized system, $\langle p_\textrm{ext}\rangle$ decays exponentially with the length:
\begin{equation}
    \langle p_\textrm{ext}(L,\cos\theta)\rangle \propto e^{-L / \lambda'},
\end{equation}
where $\lambda'$ is the (dimensional) localization length.
To analyze the finite-size scaling, we consider a dimensionless version of the localization length, defined as:
\begin{equation}
    \lambda(L, \cos\theta) \equiv -\frac{1}{\ln \langle p_\textrm{ext}(L, \cos\theta) \rangle}.
    \label{eq:lambda_kj}
\end{equation}

\section{Finite-size scaling analysis}
\label{app:size-scaling}
To extract the critical exponent associated with the localization transition, we perform a finite-size scaling analysis based on the MacKinnon–Kramer (MK) fitting procedure~\cite{MacKinnon1983}. 
In this approach, one searches for a universal scaling function that collapses localization data for different system sizes and control parameters onto a single curve.
Here, the control parameter is the out-of-plane projection of the magnetization angle $\cos\theta$ rather than a conventional disorder strength used in the original MK procedure.
The MK procedure aims to retrieve an effective size $L_\mathrm{eff}$ as a function of $\cos\theta$ such that the dimensionless localization length $\lambda(L, \cos\theta)$, given by Eq.~\ref{eq:lambda_kj}, follows single-parameter scaling:  
\begin{equation}
    \lambda(L, \cos\theta) = f\left(\frac{L}{L_\mathrm{eff}(\cos\theta)}\right),
    \label{eq:lambda_L_cosTheta}
\end{equation}
where $f$ is a universal function expected to be independent of microscopic details.
We use the terminology \emph{effective size} instead of correlation length (commonly used in the context of phase transitions) to avoid confusion with the correlation length of the disorder potential $\Lambda$.
Close to the transition, the effective size diverges as a power law:
\begin{equation} \label{eq:effective-size-power-law}
    L_\mathrm{eff}(\cos\theta)=A |\delta\!\cos\theta|^{-\nu},
\end{equation}
where $\nu$ is the critical exponent, $A$ is an undetermined coefficient and $\delta\!\cos\theta = \cos\theta - \cos\theta^*$ is the control parameter relative to the critical point.
The MK procedure obtains $L_\mathrm{eff}(\cos\theta)$ by minimizing a variance-like quantity $S$, which measures the deviation of all data points from the assumed relation for the effective size:

\begin{widetext}
\begin{equation}
\begin{split}
    S[L_\mathrm{eff}(\cos\theta)]
    &\equiv \frac{1}{N_\lambda}\sum_{i=1}^{N_\lambda} \left[
        \frac{1}{N_i} \sum_{j}
        \left[\ln L_{ij} - \ln L_\mathrm{eff}(\cos\theta_j)\right]^2
        -
        \left(
            \frac{1}{N_i} \sum_{j}
            \left[\ln L_{ij} - \ln L_\mathrm{eff}(\cos\theta_j)\right]
        \right)^2
    \right] \\[1em]
    &\sim \left\langle
        \left\langle
            [\,\ln L(\lambda,\cos\theta) - \ln L_{\mathrm{eff}}(\cos\theta)\,]^2
            \right\rangle_{\!\cos\theta}
            -
            \left\langle
            \ln L(\lambda,\cos\theta) - \ln L_{\mathrm{eff}}(\cos\theta)
        \right\rangle_{\!\cos\theta}^{2}
    \right\rangle_{\!\ln\lambda}.\label{eq:S_of_L_eff}
\end{split}
\end{equation}
\end{widetext}
On the first line, a discretized version of the variance is presented, following the original MK procedure (up to a factor $1/N_\lambda$, which does not affect the procedure). 
Here, $i$ labels different localization lengths $\lambda_i$ (not appearing explicitly in $S$), and $j$ runs over $N_i$ different control parameters $\cos\theta_j$ associated to that $\lambda_i$.
For each data point labeled by the pair of indices $(i,j)$, there is a corresponding system size $L_{ij}$ satisfying $\lambda_i=\lambda(L_{ij}, \cos\theta_j)$ (see Eq.~\ref{eq:lambda_kj}).
Minimizing $S$ yields the optimal discretized form of $L_\mathrm{eff}(\cos\theta)$ that collapses the data, from which the scaling relation of Eq.~\ref{eq:lambda_L_cosTheta} can be retrieved.
In the original MK approach, the minimization of Eq.~\ref{eq:S_of_L_eff} is achieved by solving a linear system, $\partial S / \partial \cos\theta_k = 0$.
As our data are numerically obtained for a number of different system sizes $L$ (rather than localization lengths $\lambda$), each associated with an adaptively chosen set of control parameters $\{\cos\theta_j : L\}$, we cannot directly solve this linear system.
Some form of interpolation is required to obtain a data set in the appropriate format.
Here, we consider a variation of the original approach and directly solve for the scaling exponent $\nu$ by inserting an analytical function $L(\lambda, \cos\theta)$ in Eq.~\ref{eq:S_of_L_eff}. This function is obtained via linear regression on our data points (see Appendix~\ref{app:transition-curves}) and allows us to evaluate $S$ without discretization.
Further inserting the power law of Eq.~\ref{eq:effective-size-power-law}, we have $\ln L - \ln L_{\mathrm{eff}} = \ln L + \nu \ln|\delta\!\cos\theta| - \ln A$. Since $\ln A$ is a constant irrelevant for the minimization, an alternative for Eq.~\ref{eq:S_of_L_eff} can be defined as
\begin{equation}
\begin{split}
    S(\nu) &\equiv
    \Big\langle
    \left\langle
        [\,\ln L(\lambda,\cos\theta) + \nu \ln|\delta\!\cos\theta|\,]^2
    \right\rangle_{\!\cos\theta}
    \\[0.5em]
    &\hspace{1em}
    - \left.
    \left\langle
        \ln L(\lambda,\cos\theta) + \nu \ln|\delta\!\cos\theta|
    \right\rangle_{\!\cos\theta}^{2}
    \right\rangle_{\!\ln\lambda}.
    \label{eq:S_of_nu}
\end{split}
\end{equation}
Thus, the scaling exponent $\nu$ can be directly extracted by a single-parameter minimization of Eq.~\ref{eq:S_of_nu}, with $S(\nu)$ obtained through continuous integrations over a range of $\cos\theta$ and $\ln\lambda$ values (see Appendix~\ref{app:transition-curves}).
\section{Analytical transition profile}
\label{app:transition-curves}
Observing the transition profile shown in the upper panel of Fig.~\ref{fig:1}e, we use a continuous $\tanh$ function to describe the data points:
\begin{equation} \label{eq:tanh-profile}
    \sigma_{xy}(\cos\theta) = \frac{1}{2}\left[1 + \tanh\!\left(\frac{\cos\theta - a}{b}\right)\right],
\end{equation}
where $a$ and $b$ are two independent fitting parameters.
The $L$ dependency of $\sigma_{xy}$ is now absorbed by the fitting parameters $a(L)$ and $b(L)$.
Using Eq.~\ref{eq:sigma_xy-p}, we can express $\langle p_\textrm{ext}\rangle$ in terms of the two fitting parameters:
\begin{equation}
    \langle p_\textrm{ext}\rangle
    = \frac{1}{2}\left[
        1 + 
        \frac{
            1 - \sqrt{1 - \tanh^2\!\left(\frac{\cos\theta - a}{b}\right)}
        }{
            \tanh\!\left(\frac{\cos\theta - a}{b}\right)
        }
    \right].
    \label{eq:lambda-costheta}
\end{equation}
Plugging Eq.~\ref{eq:lambda-costheta} into Eq.~\ref{eq:lambda_kj} and then taking the derivative of $\ln\lambda$ at the limit $\cos\theta\rightarrow a$, we have
\begin{figure}
    \begin{centering}
    \includegraphics[width=.9\linewidth]{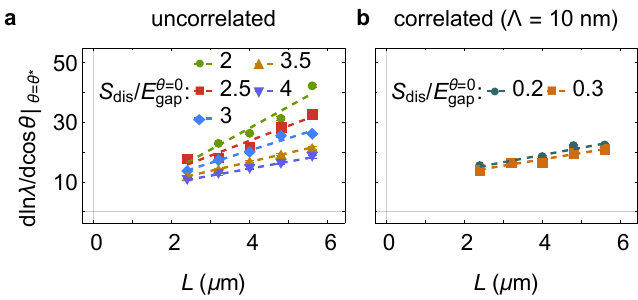}
    \end{centering}
    \caption{
        The slope $\left.\mathrm{d} \ln\lambda/\mathrm{d} \cos\theta\right|_{\theta=\theta^\ast}$ as a function of MTI ribbon length $L$ with \textbf{(a)} spatially uncorrelated and \textbf{(b)} spatially correlated ($\Lambda = 10\,\text{nm}$) disorder for different disorder strengths $S_\mathrm{dis}$, considering an $x\text{-}z$ rotation plane and Fermi level $E_\mathrm{F} = 0.2 \, E_\mathrm{gap}^{\theta=0}$. A linear fit is shown with dashed lines.
    }
    \label{fig:S2}
\end{figure}
\begin{figure}
    \begin{centering}
    \includegraphics[width=.66\columnwidth]{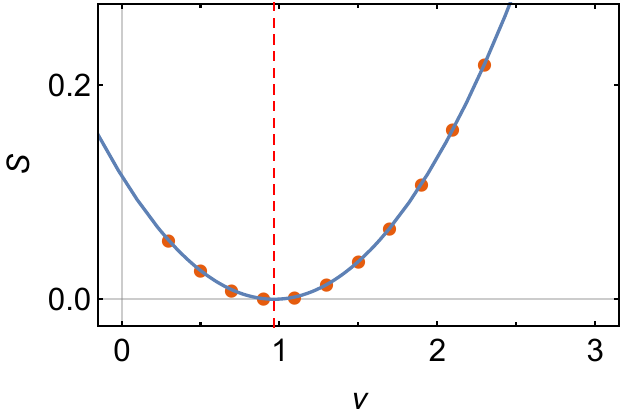}
    \end{centering}
    \caption{
        The quantity $S$ (Eq.~\ref{eq:S_of_nu}) evaluated for different values of $\nu$ for MTI ribbons with $E_\mathrm{F} = 0.2\,E_\mathrm{gap}^{\theta=0}$ and spatially uncorrelated disorder with $S_\mathrm{dis} = 2 \, E_\mathrm{gap}^{\theta=0}$, considering an $x\text{-}z$ magnetization rotation plane (see corresponding localization lengths in Figs.~\ref{fig:4}a-b). The value of $\nu$ for which $S$ is minimized, obtained via quadratic regression (blue curve), is indicated with a vertical red dashed line.
    }
    \label{fig:S3}
\end{figure}
\begin{figure}
    \centering
    \includegraphics[width=\linewidth]{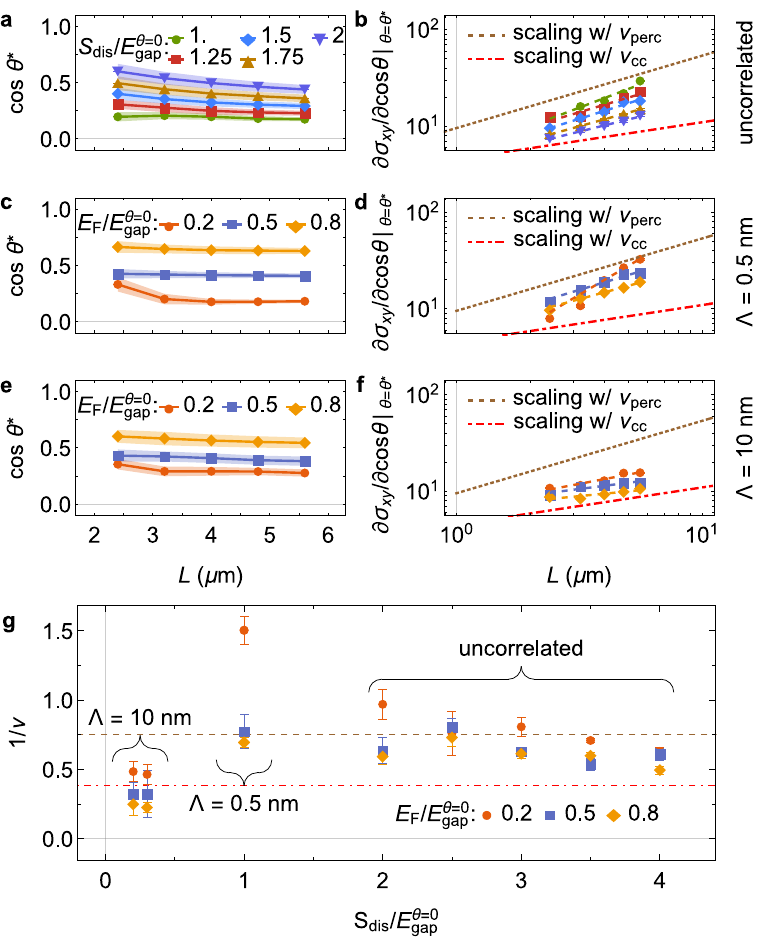}
    \caption{
        \textbf{Finite-size scaling analysis based on transition profiles.}
        \textbf{(a)-(f)} The \textbf{(a),(c),(e)} transition value $\cos\theta^\ast$ and \textbf{(b),(d),(f)} corresponding slope $\partial\sigma_{x,y}/\partial\cos\theta|_{\theta=\theta^\ast}$ as a function of MTI ribbon length $L$ for \textbf{(a),(b)} uncorrelated disorder with different disorder strengths and $E_\mathrm{F} = 0.2 \, E_\mathrm{gap}^{\theta=0}$, and for \textbf{(c)-(f)} correlated disorder with different Fermi levels, considering \textbf{(c),(d)} $\Lambda = 0.5\,\text{nm}$, $S_\mathrm{dis} = E_\mathrm{gap}^{\theta=0}$ and \textbf{(e),(f)} $\Lambda = 10\,\text{nm}$, $S_\mathrm{dis} = 0.2 \, E_\mathrm{gap}^{\theta=0}$.
        \textbf{(g)} The extracted values for $1/\nu$ as a function of disorder strength, for uncorrelated and correlated disorder with different correlation lengths, and different Fermi level positions.
        The classical ($\nu_\mathrm{perc} = 4/3$) and quantum ($\nu_\textsc{cc} = 2.6$) percolation exponents are shown for reference [with corresponding scaling curves in \textbf{(b),(d),(f)}].
        For these results, the magnetization is rotated in the $x\text{-}z$ plane.
    }
    \label{fig:S4}
\end{figure}
\begin{equation}
    \left. \frac{\mathrm{d} \ln \lambda}{\mathrm{d} \cos\theta} \right|_{\cos\theta = a}
    = \frac{1}{b \ln 4},
\end{equation}
where $b$ is a function of $L$, containing the scaling behavior near the critical point.
Due to the numerical complexity of the transport simulation, the range of $L$ is small.
We can therefore linearize the $L$ dependency as
\begin{equation} \label{eq:b-inv}
    1/b \approx c_0 + c_1 L,
\end{equation}
where $c_0$ and $c_1$ are coefficients only depending on the system parameters including the Fermi level and the disorder characteristics.
These two coefficients can be obtained via linear regression of the discrete points $1/b_k(L_k)$, and the result can be see in Fig.~\ref{fig:S2}.
Thus, by combining Eqs.~\ref{eq:lambda_kj}, \ref{eq:lambda-costheta}, \ref{eq:b-inv}, and letting $\delta\!\cos\theta = \cos\theta - a$, we obtain an analytical expression for $L(\lambda, \cos\theta)$:
\begin{equation}
    L(\lambda,\cos\theta) =
    \frac{
        \operatorname{arctanh}\!\left[
        \dfrac{e^{1/\lambda}(2 - e^{1/\lambda})}{
        2 - 2e^{1/\lambda} + e^{2/\lambda}}
        \right]
        - c_0\,\delta\!\cos\theta
    }{
        c_1\,\delta\!\cos\theta
    }.
    \label{eq:L_lambda_compact}
\end{equation}
Plugging Eq.~\ref{eq:L_lambda_compact} in Eq.~\ref{eq:S_of_nu}, we can perform a single-parameter minimization to obtain the scaling exponent $\nu$.
The averages $\langle\cdots\rangle$ in Eq.~\ref{eq:S_of_L_eff} run over bounded intervals of $\delta\!\cos\theta$ and $\ln\lambda$.
For our analysis presented in Fig.~\ref{fig:4}, we have $L_\mathrm{min} = 2.4\,\textnormal{\textmu m}$, $L_\mathrm{max} = 5.6\,\textnormal{\textmu m}$, $\ln \lambda_\mathrm{min} = \ln\lambda_\mathrm{c} - 1$, and $\ln \lambda_\mathrm{max} = \ln\lambda_\mathrm{c} + 1.5$, where $\lambda_\mathrm{c}=1/\ln2$ is the critical value according to Eq.~\ref{eq:lambda_L_cosTheta}.
In practice, the value of $\nu$ is obtained from a quadratic regression, an example of which is shown in Fig.~\ref{fig:S3}.
This indeed collapses all the simulated data points, as illustrated in Fig.~\ref{fig:4}b, considering Eq.~\ref{eq:effective-size-power-law} with $A=1\,\text{\textmu m}$, thus confirming the single-parameter relation of Eq.~\ref{eq:lambda_L_cosTheta}.
\section{Extracting $\nu$ directly from $\sigma_{xy}$}
\label{app:size-scaling-alt}
To examine the numerical accuracy of our single-parameter optimization shown in Eq.~\ref{eq:S_of_nu}, here we extract the critical exponent in an alternative way and compare.
According to finite-size scaling theory~\cite{Pruisken1988, Huckestein1995}, the (Hall) conductivity displays single-parameter scaling of the form:
\begin{equation}
    \sigma_{xy}(L, \cos\theta) = \frac{1}{2}\frac{e^{2}}{h}g\left(L^{\frac{1}{\nu}} \, \delta\!\cos\theta \right).
    \label{eq:sigma_xy_powerLaw}
\end{equation}
where $g$ is a universal function similar to $f$ in Eq.~\ref{eq:effective-size-power-law}.
Instead of performing the single-parameter optimization, a finite-size scaling analysis can be directly performed by fitting the maximal transition slope $\partial \sigma_{xy} / \partial \cos\theta|_{\theta=\theta_\mathrm{c}}$ as a function of system size $L$ to a power law~\cite{Kawamura2020}.
Taking the derivative and natural logarithm on both sides of Eq.~\ref{eq:sigma_xy_powerLaw}, we have
\begin{equation}
    \ln\left( \left. \frac{\partial\sigma_{xy}(L, \cos\theta)}{\partial \cos\theta} \right|_{\theta = \theta_\mathrm{c}} \right) = \frac{1}{\nu} \ln L + \ln\left(\frac{1}{2}\frac{e^2}{h}g'(0) \right),
    \label{eq:linearFitting}
\end{equation}
where the exponent can be extracted from the slope in a linear regression.
These slopes and the related results are presented in Fig.~\ref{fig:S4}.
Specifically, the slopes in Figs.~\ref{fig:S4}b, \ref{fig:S4}d and \ref{fig:S4}f agree well with our single-parameter optimization results presented in Fig.~\ref{fig:4}c.
Besides serving as a numerical stability check, Eq.~\ref{eq:linearFitting} also provides a convenient way to extract the finite-size scaling exponent
$\nu$ directly from experimental data.
%

%\bibliographystyle{apsrev4-2}
\bibliography{references}

\end{document}